\documentclass[a4paper]{article}
\usepackage{xcolor}
\usepackage[utf8]{inputenc}
\usepackage[T1]{fontenc}
\usepackage[polish,english]{babel}
\usepackage{graphicx}\graphicspath{{img/}}
\usepackage{lmodern,fancyhdr,secdot,float,amsmath,amsfonts,amsthm,amssymb}
\usepackage{booktabs,tabularx}
\usepackage[margin=2cm]{geometry}
\usepackage{multicol}
\usepackage[hidelinks,unicode]{hyperref}
\usepackage{titlesec}
\usepackage{indentfirst}
\titleformat{\section}{\centering\normalsize\bf}{\thesection.}{.5em}{\MakeUppercase}
\titleformat*{\subsection}{\bf\normalsize\selectfont}
\titleformat*{\subsubsection}{\bf\normalsize\selectfont}
\newcommand{\titlePL}[1]{\large\textbf{ #1}}
\newcommand{\titleEN}[1]{\normalsize #1}
\newcommand{\keywordsPL}[1]{\small\textbf{Słowa kluczowe:} #1}
\newcommand{\keywordsEN}[1]{\small\textbf{Keywords:} #1}
\newcommand{\abstractPL}[1]{\small\textbf{Streszczenie:} #1}
\newcommand{\abstractEN}[1]{\small\textbf{Abstract:} #1}

\definecolor{logo_color}{RGB}{40, 69, 166}

\begin{document}\thispagestyle{empty}\pagestyle{fancy}
\begin{minipage}[t]{0.5\textwidth}\vspace{0pt}%
\includegraphics[width=5.22cm]{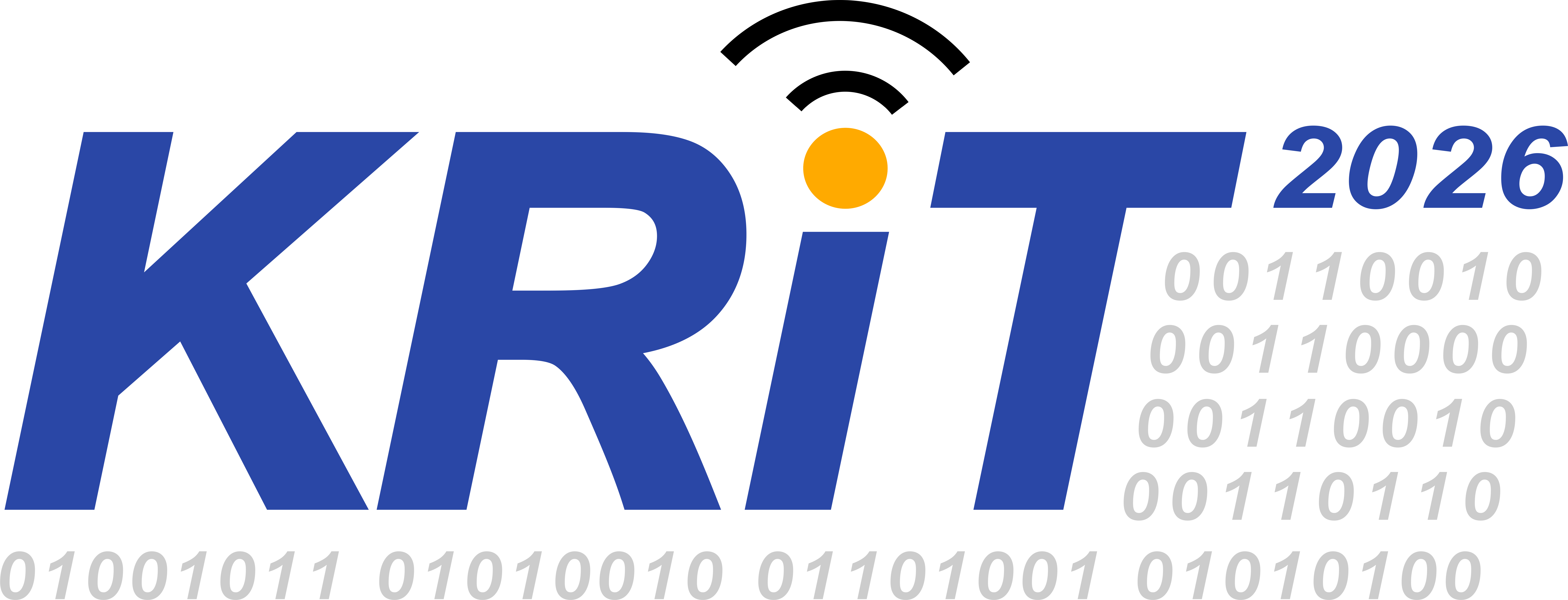}
\end{minipage}
\begin{minipage}[t]{0.45\textwidth}\vspace{12pt}%
\centering
\color{logo_color} KONFERENCJA RADIOKOMUNIKACJI\\ I TELEINFORMATYKI\\ KRiT 2026
\end{minipage}

\vspace{1cm}

\begin{center}
\titlePL{Agent AI Zarządzający Siecią Open RAN Zorientowana na Użytkownika}

\titleEN{Agentic AI for Management of Open RAN User-Centric Cell-Free Network}\medskip

Marcin Hoffmann$^{1,2}$

\medskip

\begin{minipage}[t]{0.6\textwidth}
\small $^{1}$ Politechnika Poznańska, Poznań, \href{mailto:email}{marcin.hoffmann@put.poznan.pl} \\
\small $^{2}$ Rimedo Labs, Poznań \\
\end{minipage}

\medskip
\end{center}

\medskip

\begin{multicols}{2}
\noindent
\abstractPL{
Jednym z najważniejszych wyzwań związanych z implementacją sieci Open RAN o architekturze zorientowanej na użytkownika jest odpowiednie formowanie tzw. klastrów obsługujących. Problem ten może być efektywnie rozwiązany za pomocą zaproponowanego Agenta AI bazującego na modelu językowym. Wyniki symulacji pokazały, że zaproponowane rozwiązanie pozwala na ponad 2-krotne zwiększenie mediany rozkładu przepływności użytkowników względem standardowej architektury zorientowanej na sieć.\footnote[1]{Praca powstała w~ramach projektu PRELUDIUM finansowanego przez Narodowe Centrum Nauki nr 2022/45/N/ST7/01930.}}
\medskip

\noindent
\abstractEN{
One of the main challenges associated with the implementation of an Open RAN User-Centric Cell-Free network is the appropriate formulation of serving clusters. This problem can be effectively solved using a proposed AI Agent, based on a large language model. Simulation results have shown that the proposed solution allows for over a 2-fold increase in the median of the user throughput distribution compared to the state-of-the-art network-centric architecture.}
\medskip

\noindent
\keywordsPL{AI, 6G, architektura zorientowana na użytkownika, Massive MIMO, Duże Modele Językowe}
\medskip

\noindent
\keywordsEN{AI, 6G, User-Centric Cell-Free Architecture, Massive MIMO, Large Language Models.}

\section{Wstęp}

Jedną z kluczowych technik pozwalających znacznie poprawić efektywność widmową sieci mobilnych szóstej generacji (6G) jest zastosowanie tzw. architektury zorientowanej na użytkownika (\textit{ang. User-Centric Cell-Free})~\cite{yang2021}. W ramach nowej architektury użytkownik może być jednocześnie obsługiwany przez wiele zsynchronizowanych stacji bazowych (nazywanych też często punktami dostępowymi), najczęściej wyposażonych w macierze antenowe M-MIMO (\textit{ang. Massive Multiple-Input Multiple-Output}). Nowa architektura pozwala na zmniejszenie dysproporcji w mocy odbieranej w różnych punktach sieci. Jest to szczególnie istotne w przypadku użytkowników znajdujących się w podobnej odległości od kilku stacji bazowych~\cite{ngo2015}. Ponadto możliwe jest skoordynowane sterowanie wiązką w celu zmniejszenia interferencji. Z punktu widzenia praktycznego wdrożenia architektury zorientowanej na użytkownika duży potencjał ma koncepcja otwartej sieci dostępowej Open RAN (\textit{ang. Open Radio Access Network})~\cite{Ranjbar2022,Hoffmann_2024_infocom}. Zakłada ona tzw. podział 7.2, pozwalający na rozdzielenie funkcjonalności stacji bazowej na zwirtualizowaną jednostkę O-CU (\textit{Open RAN Centralized Unit}) odpowiedzialną m.in. za zarządzanie zasobami radiowymi, zwirtualizowaną jednostkę O-DU (\textit{Open RAN Distributed Unit}), realizującą funkcje warstwy fizycznej wyższego poziomu i warstwy dostępu do łącza, oraz fizyczny O-RU (\textit{Open RAN Radio Unit}), odpowiedzialny np. za funkcje warstwy fizycznej niższego poziomu i realizujący transmisję radiową. Z perspektywy sieci zorientowanej na użytkownika jedno O-CU i jedno O-DU zaimplementowane w wirtualnym środowisku mogą koordynować transmisję z wielu O-RU, pełniąc funkcję CPU (\textit{ang. Central Processing Unit}). Dodatkowo koncepcja Open RAN poprzez zastosowanie otwartych interfejsów i jednostki RIC (\textit{ang. RAN Intelligent Controller}) pozwala na implementację dedykowanych algorytmów optymalizujących sieć, np. z wykorzystaniem metod AI (\textit{ang. Artificial Intelligence}), w tym agentów AI bazujących na modelach językowych~\cite{Wu2025}. 

Ze względu na ograniczone zasoby obliczeniowe po stronie O-CU/O-DU oraz O-RU, jednym z głównych wyzwań związanych  z zarządzaniem siecią zorientowaną na użytkownika jest tzw. formowanie klastrów obsługujących~\cite{ammar2022}. Procedura ta ma na celu określenie zbioru O-RU, które powinny obsługiwać konkretnego użytkownika. Ze względu na dużą liczbę możliwych kombinacji jest to zadanie nietrywialne. Ponadto wykonane niewłaściwie może wręcz pogorszyć wydajność sieci, np. w przypadku gdy zbyt duża liczba O-RU obsługuje użytkownika, część znacząco oddalonych O-RU będzie miała marginalny wpływ na poprawę jego przepływności, "marnując" w ten sposób cenne zasoby radiowe. Z drugiej strony, zbyt mała liczba O-RU nie zapewni wzrostu odbieranej mocy i koordynacji interferencji. Z tej perspektywy rozsądne wydaje się zastosowanie wspomnianych metod AI, ze szczególnym uwzględnieniem modeli językowych, które pozwolą na interakcję ze środowiskiem i interpretację wyników podobną do procesu konfiguracji parametrów sieci dokonywanego przez człowieka, jednak odbywająca się w sposób ciągły i automatyczny. 

W tej pracy proponuję Agenta AI zarządzającego siecią Open RAN zorientowaną na użytkownika. W związku z bardzo dużą liczbą możliwych kombinacji przypisania użytkowników do O-RU proponuję podejście hierarchiczne, w którym Agent AI zaimplementowany jako rApp modyfikuje parametry algorytmu formowania klastrów obsługujących zaimplementowanego bezpośrednio w O-CU/O-DU w zależności od lokalizacji użytkowników w sieci. Zaproponowany Agent AI maksymalizuje medianę rozkładu przepływności użytkowników. Dodatkowo w odróżnieniu od metod MAB (\textit{ang. Multi-Arm Bandit}) wykorzystuje historyczną wiedzę, zmniejszając czas potrzebny na odkrywanie nowych rozwiązań. Zaproponowany Agent AI jest zweryfikowany z użyciem zaawansowanego symulatora sieci zorientowanej na użytkownika wykorzystującego trójwymiarowe śledzenie promieni (\textit{ang. Ray-Tracing}), przydział zasobów radiowych w czasie-częstotliwości-przestrzeni, schematy modulacji i kodowania MCS (\textit{ang. Modulation and Coding Schemes}) i prekoder ZF (\textit{Zero-Forcing})~\cite{hoffmann2024}. W dalszej części tej pracy w rozdziale~\ref{sec:ai_agent} opisany jest zaproponowany Agent AI. W Rozdziale~\ref{sec:results} przedstawione są wyniki, a wnioski są sformułowane w Rozdziale~\ref{sec:conclusions}.

\section{Agent AI Zarządzający Siecią Zorientowaną na Użytkownika} \label{sec:ai_agent}
\begin{figure}[H]
\centering
\includegraphics[trim={11cm 1.5cm 11cm 1.5cm},clip,width=0.49\textwidth]{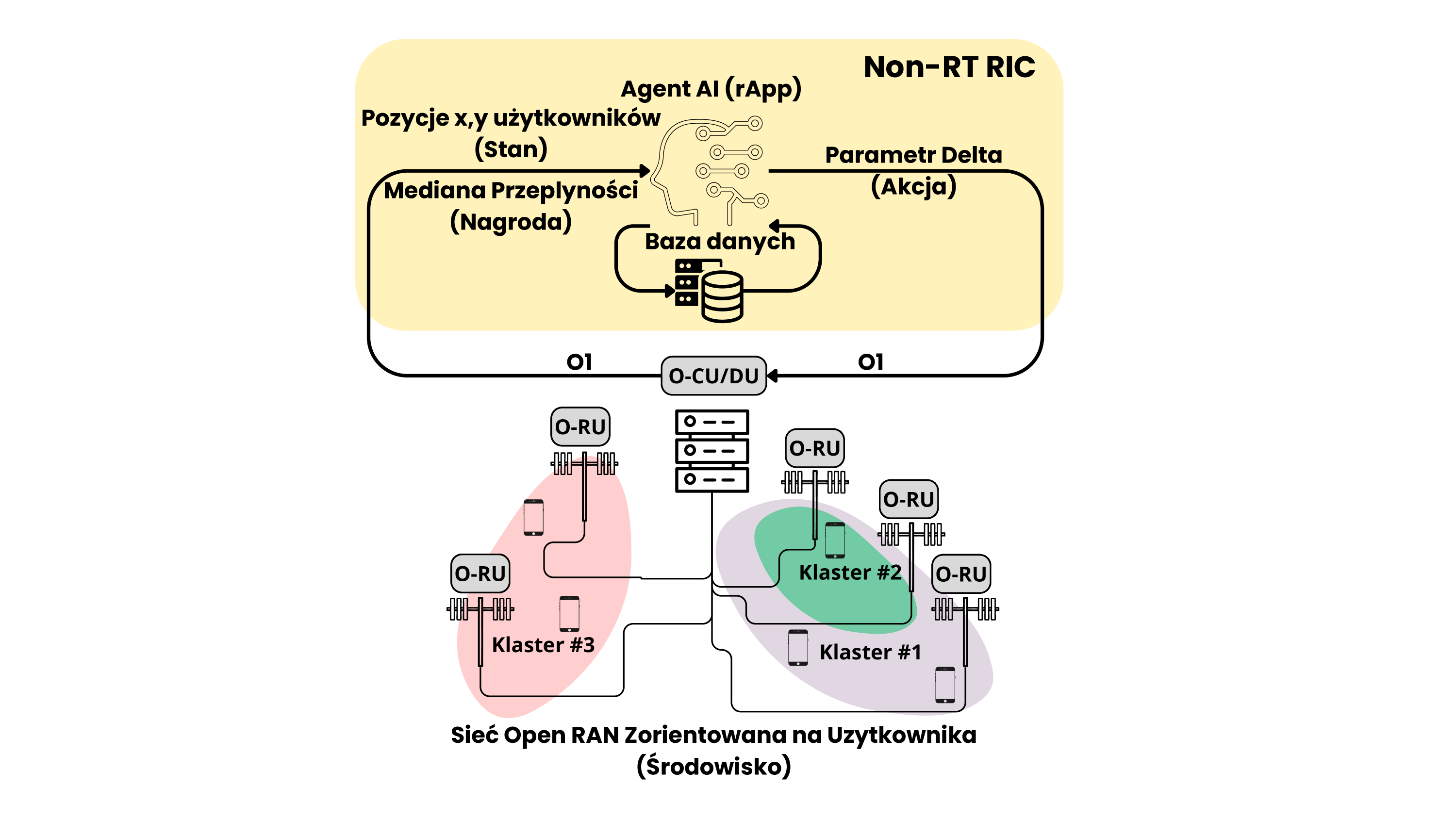}
\caption{Agenta AI zarządzający formowaniem klastrów obsługujących w sieci zorientowanej na użytkownika.}
\label{fig:system}
\end{figure}
W tej pracy proponuję Agenta AI odpowiedzialnego za formowanie klastrów obsługujących w łączu w dół w sieci Open RAN zorientowanej na użytkownika, jak przedstawiono na Rys.~\ref{fig:system}. W tej sieci wiele O-RU wyposażonych w macierze M-MIMO jest podłączonych do zwirtualizowanego O-CU/O-DU pełniącego rolę CPU. Poza przydziałem zasobów radiowych oraz koordynacją transmisji do użytkowników z fizycznych jednostek radiowych O-RU, jednym z najistotniejszych zadań O-CU/O-DU jest zestawienie klastrów obsługujących. Wymaga ono odpowiedniej konfiguracji O-RU tworzących ten klaster w celu zapewnienia m.in. wymiany danych użytkownika, synchronizacji transmisji oraz wymiany informacji o współczynnikach kanałów radiowych poszczególnych użytkowników. W szczególności potencjalnym wyzwaniem jest formowanie klastrów obsługujących wymagających koordynacji pomiędzy dwoma O-CU/O-DU, tj. fragmentami sieci obsługiwanymi przez różne CPU. Zanim jednak to nastąpi należy wskazać którzy użytkownicy powinni być obsługiwani przez które O-RU - dokonać formowania klastrów obsługujących. W ramach zaproponowanego schematu zarządzania założono, że O-CU/O-DU posiada wbudowany algorytm formowania klastrów obsługujących o niskiej złożoności tj., odpowiedni do zastosowań wymagających czasu reakcji na poziomie pojedynczych milisekund. 
Taki algorytm został zaproponowany w~\cite{tugfe2021cellfree}. Klaster obsługujący $k$-tego użytkownika (oznaczony symbolem $\mathcal{M}_k$) tworzony jest przez minimalną liczbę O-RU $N_k$, wybranych spośród wszystkich $L$ O-RU (indeksowanych przez zmienną~$l$), o największej mocy nadawanej do użytkownika $p_{k,l}$ spełniających warunek:
\begin{equation}\label{eq:serving_cluster_formulation}
\frac{\sum_{l \in \mathcal{M}_k}p_{k,l}}{\sum_{l=1}^{\mathrm{L}}p_{k,l}} \geq \Delta,
\end{equation}
gdzie $\Delta$ jest parametrem algorytmu w zakresie od $0$ do $1$. Wartości $\Delta$ bliskie $0$ odpowiadają architekturze zorientowanej na sieć (jeden użytkownik obsługiwany przez 1 O-RU o największej mocy odbieranej), a wartość $\Delta=1$ oznacza, że użytkownik jest obsługiwany przez wszystkie O-RU w swoim zasięgu. Z tej perspektywy, głównym wyzwaniem jest odpowiednie dobranie parametru $\Delta$ w zależności od aktualnego rozmieszczenia użytkowników w sieci.

W tym celu zaproponowałem Agenta AI, który w związku z relatywnie długim czasem odpowiedzi (rzędu kilku sekund) i dużymi wymaganiami na zasoby obliczeniowe jest zaimplementowany jako rApp w Non-RT RIC (patrz Rys.~\ref{fig:system}). Interakcja Agenta AI ze środowiskiem bazuje na cyklu uczenia ze wzmocnieniem i wykorzystuje interfejs O1 do pobierania danych o sieci oraz zmiany parametru $\Delta$, tzn. Agent AI uczy się poprzez interakcję ze środowiskiem, tzn. cyklicznie rozpoznaje stan, wykonuje akcję, korzystając z historycznej wiedzy zapisanej w bazie danych, i obserwuje nagrodę. Poszczególne elementy cyklu zdefiniowane są w sposób następujący:
\begin{itemize}
    \item \textbf{Stan} reprezentowany jest przez macierz pozycji użytkowników wyrażonych jako współrzędne X,Y.
    \item \textbf{Akcja} polega na wyborze przez agenta wartości parametru $\Delta$ z zakresu od 0 do 1.
    \item \textbf{Nagroda} jest zdefiniowana jako mediana rozkładu przepływności użytkowników.
\end{itemize}
Zaproponowany Agent AI bazuje na modelu językowym, a sposób w jaki wybiera akcję na bazie aktualnego stanu i danych historycznych zdefiniowany jest w sposób tekstowy w ramach tzw.
\begin{itemize}
    \item \textbf{System Prompt}, który określa kontekst sytuacyjny oraz ogólny sposób zachowania Agenta AI. W zaproponowanym przeze mnie system prompt informuję agenta, że jest odpowiedzialny za zarządzanie siecią zorientowaną na użytkownika składającą się z $L$ O-RU. Następnie podaję definicję parametru $\Delta$ tzn. jego zakres i interpretację. W końcu informuję agenta, że będą mu podawane jako dane wejściowe położenia użytkowników (stan) oraz dane historyczne zawierające informację o medianie z rozkładu przepływności użytkowników zaobserwowanej dla danych położeń i parametru $\Delta$. Na tej podstawie agent powinien zwrócić wartość parametru $\Delta$ (akcję), która zapewni największą wartość mediany z rozkładu przepływności użytkowników (nagrodę). 
    \item \textbf{User Prompt}, to bezpośrednie zapytanie kierowane do Agenta AI, zawierające dane wejściowe. Dla zaproponowanego Agenta AI user prompt zawiera informację o aktualnych położeniach użytkowników w sieci (stan), oraz dane historyczne, które pozwalają agentowi wnioskować o tym jaki parametr $\Delta$ (akcję) wybrać w aktualnym stanie na podstawie otrzymanych poprzednio nagród. Ponadto, agent używając danych historycznych może wnioskować o tym jak zachować się w nowym do tej pory nie odwiedzonym stanie, bez konieczności potarzania procesu uczenia niezależnie dla różnych położeń użytkowników: stanów.  
\end{itemize}

\section{Wyniki Symulacji} \label{sec:results}
\begin{figure}[H]
\centering
\includegraphics[width=0.49\textwidth]{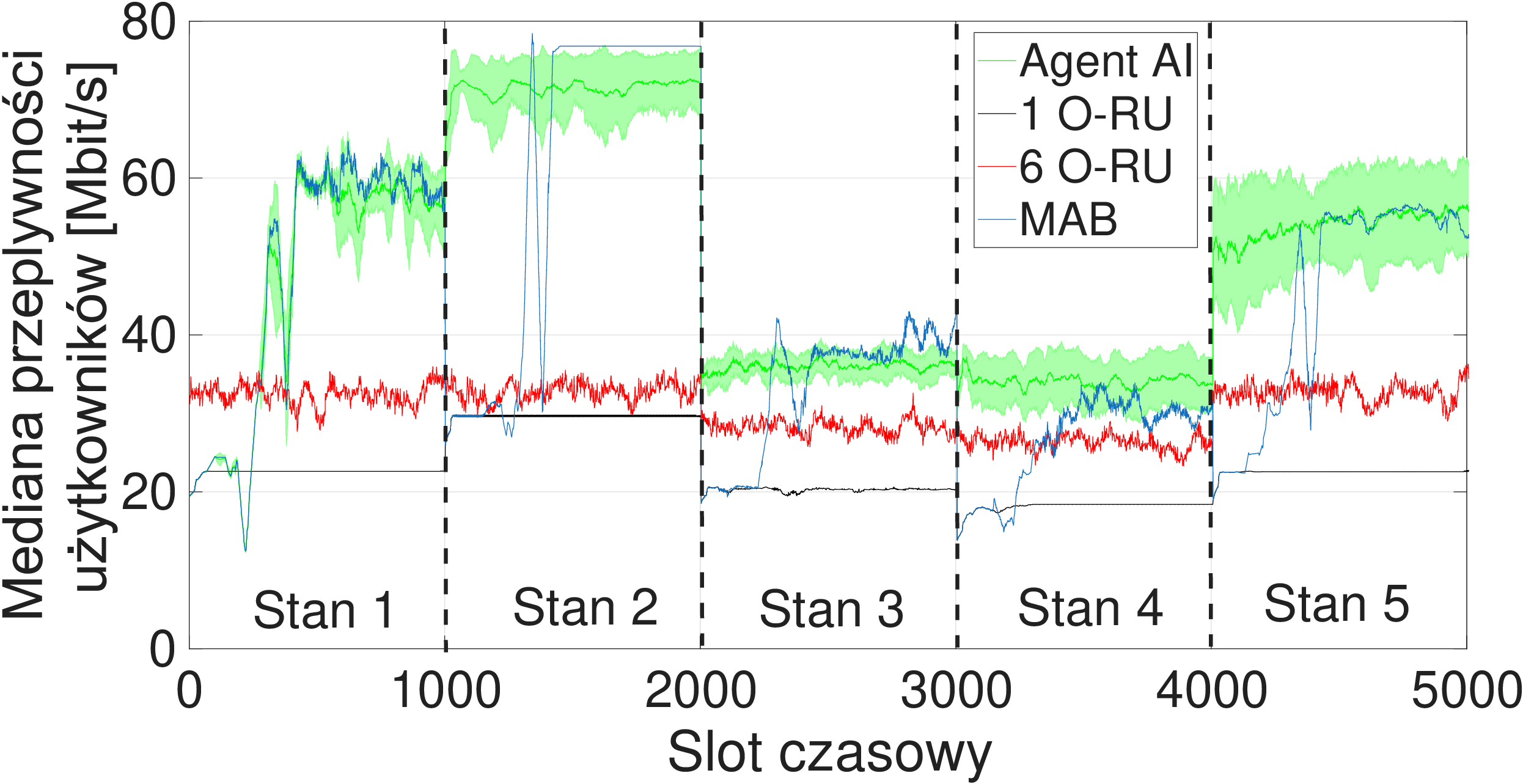}
\caption{Mediana przepływności użytkowników w funkcji czasu. Dla Agenta AI uwzględniono średnią i odchylenie standardowe z 10 niezależnych przebiegów symulacyjnych.}
\label{fig:learning_time}
\end{figure}

Zaproponowany Agent AI zarządzający formowaniem klastrów obsługujących w sieci Open RAN zorientowanej na użytkownika został zbadany za pomocą zaawansowanego symulatora zaproponowanego przez autora w poprzedniej pracy~\cite{hoffmann2024}. Symulator modeluje sieć M-MIMO o architekturze zorientowanej na użytkownika i wykorzystującej OFDMA. Zastosowano w nim realistyczny model kanału radiowego oparty na trójwymiarowym śledzeniu promieni firmy Wireless InSite, jak również podział na wiele bloków funkcjonalnych, takich jak: formowanie klastrów obsługujących, dedykowany algorytm przydziału zasobów radiowych, prekoder ZF (\textit{ang. Zero-Forcing}) działający niezależnie dla każdej jednostki radiowej O-RU i 15 schematów modulacji i kodowania. W ramach jednej jednostki radiowej O-RU, podział mocy w obrębie pojedynczego bloku zasobów odbywa się równo pomiędzy wszystkich użytkowników, którym został on przydzielony w rezultacie multipleksacji przestrzennej. Na potrzeby eksperymentów symulacyjnych zaproponowany Agent AI wykorzystuje model językowy Qwen 3.5-32b zainstalowany na lokalnej stacji roboczej z wykorzystaniem oprogramowania Ollama. Pozostałe parametry przeprowadzonych symulacji zostały zestawione w Tabeli~\ref{tab1}.
\begin{table}[H]
\centering
\caption{Parametry Symulacji}
\label{tab1}
\medskip
\begin{tabularx}{1.0\columnwidth}{p{0.65\linewidth} p{0.17\linewidth} }
\toprule
\textit{Parametr} & \textit{Wartość}\\ \midrule
Model językowy	& Qwen 3.5 32b \\ \hline
Częstotliwość nośna		& 3.6~GHz \\ \hline
Liczba bloków zasobów & 69 \\ \hline
Liczba O-CU/O-DU & 1 \\ \hline
Liczba jednostek radiowych O-RU & 1 makro, 5 mikro \\ \hline
Wysokość anten - makro O-RU & 45~m\\ \hline
Wysokość anten - mikro O-RU & 6~m\\ \hline
Moc nadawana - makro O-RU & $46$~dBm \\ \hline
Moc nadawana - mikro O-RU & $30$~dBm \\ \hline
Liczba anten - makro O-RU & 128\\ \hline
Liczba anten - mikro O-RU & 32 \\ \hline
Czas symulacji (sloty) & 5000 \\ \hline
Liczba użytkowników & 20 \\ \hline
Wireless InSite - liczba odbić & 15 \\ \hline
Wireless InSite - liczba dyfrakcji & 1\\\bottomrule
\end{tabularx}
\end{table}
W eksperymencie symulacyjnym porównałem zaproponowanego Agent AI z architekturą zorientowaną na sieć (oznaczenie 1 O-RU na wykresach), architekturą zorientowaną na użytkownika, gdzie wszystkie O-RU obsługują wszystkich użytkowników (oznaczenie 6 O-RU), oraz wyborem parametru $\Delta$ na bazie algorytmu kontekstowego wielorękiego bandyty (oznaczenie MAB - \textit{ang. Multi-Arm Bandit})~\cite{Hoffmann_2024_krit}. Algorytmy zostały porównane dla 5 niezależnych stanów, tzn. niezależnie wylosowanych rozkładów położeń użytkowników. Dodatkowo, w związku z tym, że modele językowe bazują na prawdopodobieństwie i ich odpowiedzi mogą różnić się pomiędzy wywołaniami, dla zaproponowanego Agenta AI wykonałem 10 niezależnych przebiegów symulacyjnych, używając tej samej sekwencji stanów. Wynikowy wykres mediany z rozkładu przepływności użytkowników w czasie przedstawiony jest na Rys.~\ref{fig:learning_time}. Pierwszą obserwacją jest to, że w rozważanych stanach architektura zorientowana na sieć (1 O-RU) osiąga najniższe wartości mediany z rozkładu przepływności użytkowników w każdym stanie: od około 20 do około 30 Mbit/s. Z drugiej strony, skrajna konfiguracja sieci zorientowanej na użytkownika, gdzie każdy użytkownik jest obsługiwany przez wszystkie O-RU, pozwala tylko nieznacznie poprawić medianę przepływności do zakresu około 35–25 Mbit/s. Znaczący wzrost mediany z rozkładu przepływności użytkowników może być osiągnięty poprzez dobór parametru $\Delta$ pomiędzy 0 a 1. W przypadku MAB dobór parametru $\Delta$ odbywa się niezależnie w obrębie każdego stanu i pozwala osiągnąć wartości mediany z rozkładu przepustowości użytkowników z zakresu od 35 Mbit/s do nawet około 70 Mbit/s w Stanie 2. Z jednej strony, niezależny dobór w każdym stanie pozwala w pojedynczych przypadkach (np. Stan 2 i 3) znaleźć minimalnie lepszą wartość niż proponowany Agent AI; z drugiej strony, wymaga to poświęcenia około 1/3 czasu na sprawdzenie wszystkich wartości parametru $\Delta$, również takich, które pogarszają medianę przepływności. Wreszcie, widać, że zaproponowany Agent AI wykonuje pełny proces uczenia tylko  w Stanie 1, a następnie korzysta z zebranej wiedzy i dokonuje tylko drobnych korekt. W większości stanów średnia mediana przepływności osiągana przez Agenta AI jest podobna do wartości osiąganych przez MAB, jednak odbywa się to (poza Stanem 1) bez konieczności każdorazowego powtarzania procesu uczenia. Oczywiście, w zależności od realizacji mediana rozkładu przepływności użytkowników oscyluje w ramach odchylenia standardowego rzędu kilku Mbit/s. 

\begin{figure}[H]
\centering
\includegraphics[width=0.49\textwidth]{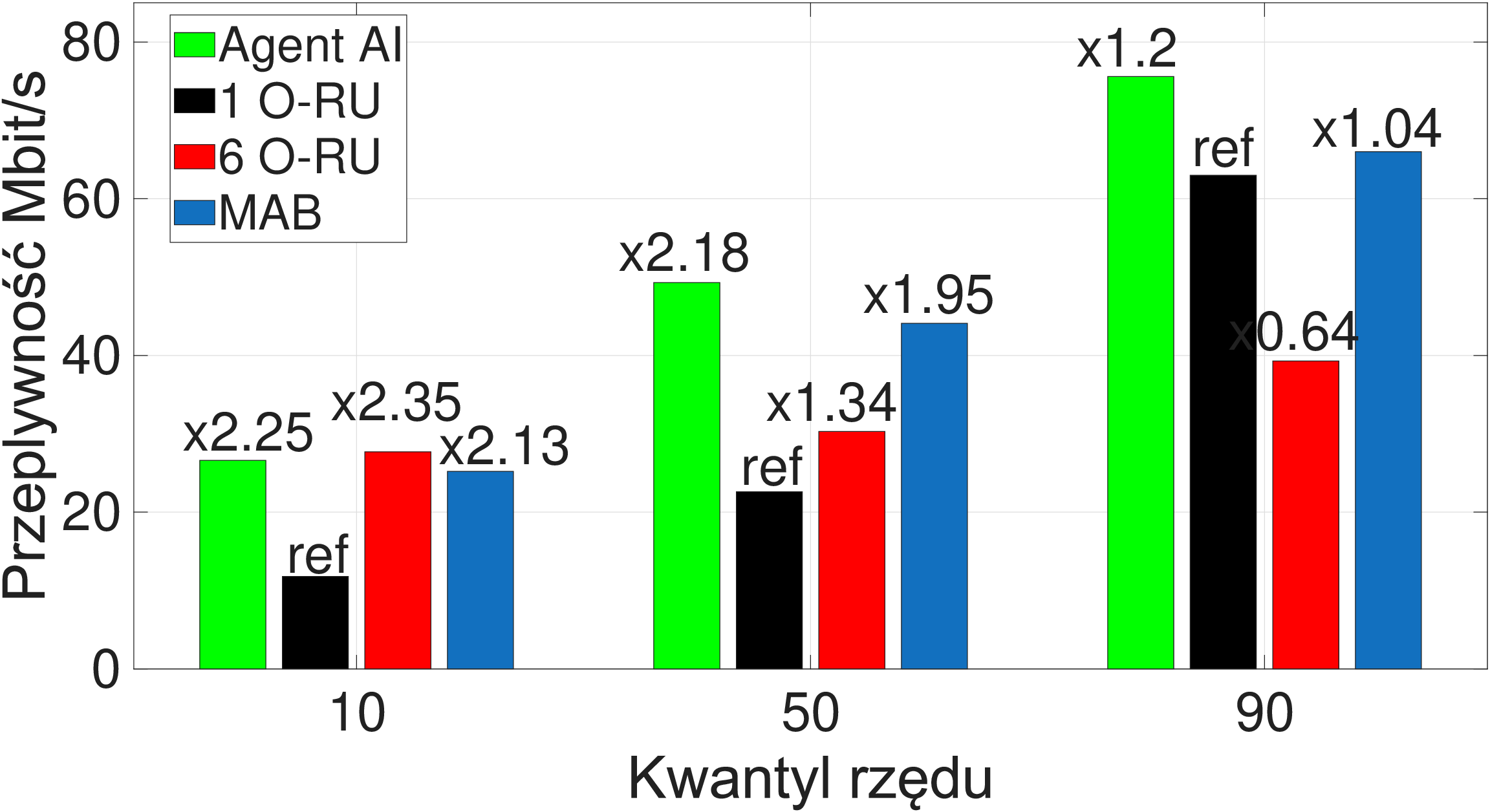}
\caption{Porównanie kwantyli rzędu 10, 50, i 90 z rozkładu przepływności użytkowników z całej symulacji.}
\label{fig:rate_bar}
\end{figure}

Z perspektywy porównania Agenta AI z MAB istotne jest porównanie mediany z rozkładu przepływności użytkowników dla całego eksperymentu symulacyjnego. Takie wyniki przedstawione są na Rys.~\ref{fig:rate_bar}, dodatkowo uwzględniając kwantyle rzędu 10 i 90, reprezentujące odpowiednio przepływność użytkowników mających najgorsze i najlepsze warunki propagacyjne. Za punkt odniesienia przy analizie służy wynik dla architektury zorientowanej na sieć (1 O-RU). Dla architektury zorientowanej na użytkownika, gdzie wszyscy użytkownicy są obsługiwani przez wszystkie O-RU, obserwujemy 1,34-krotny zysk w medianie przepływności oraz 2,35-krotny zysk dla użytkowników charakteryzujących się najgorszymi warunkami propagacyjnymi. Dla tych użytkowników koordynacja interferencji i zwiększenie mocy odbieranej są kluczowe. Niestety, odbywa się to kosztem użytkowników o najlepszych warunkach propagacyjnych, dla których osiągane przepływności to tylko 64\% referencyjnych. Może to być spowodowane koniecznością podziału mocy nadawanej na bardzo wielu, często odległych użytkowników. Z drugiej strony wyniki pokazują zalety z dynamicznego dostosowywania parametru $\Delta$, widoczne zarówno dla zaproponowanego Agenta AI, jak i MAB. W przypadku mediany z rozkładu przepływności użytkowników zaproponowany Agent AI poprawia wynik dla architektury zorientowanej na sieć 2,18 raza, a MAB tylko 1,95: zysk około 12\% na korzyść Agenta AI. Wynika to głównie ze zdolności Agenta AI do wnioskowania co do wartości parametru $\Delta$ w Stanach 2, 3, 4, 5 na podstawie szczątkowych danych historycznych, np. ze Stanu 1 bez konieczności przechodzenia każdorazowo niezależnego procesu uczenia, jak to ma miejsce w przypadku MAB. Podobne zależności widoczne są dla użytkowników o najmniej korzystnych warunkach radiowych: Agent AI zapewnia zysk 2,25-krotny, a MAB 2,13-krotny. W końcu dla użytkowników charakteryzujących się najlepszymi warunkami propagacyjnymi Agent AI zapewnia poprawę przepływności o około 20\%, w stosunku do MAB zapewniającego tylko około 4\%.

\section{Wnioski} \label{sec:conclusions}
Formowanie klastrów obsługujących jest jednym z kluczowych wyzwań dla sieci 6G zorientowanych na użytkownika. W tej pracy zaproponowałem wykorzystanie do tego celu Agenta AI bazującego na modelu językowym. W odróżnieniu od metod MAB Agent AI może wykorzystywać wnioskowanie zbliżone do ludzkiego na podstawie danych historycznych, co pozwala utrzymywać wysoką medianę rozkładu przepływności użytkowników, dokonując jedynie drobnych korekt dla zmieniających się położeń użytkowników. W rezultacie zaproponowany Agent AI poprawi medianę przepływności użytkowników o około 12\% w stosunku do rozwiązania MAB. Jednym z dalszych kierunków rozwoju jest porównanie wyników osiąganych przez zaproponowanego Agenta AI dla różnych modeli językowych, zarówno mniejszych jak i dużych komercyjnych modeli ulokowanych w chmurze np. ChatGPT lub Gemini.
\bibliographystyle{krit}
\bibliography{references}
\end{multicols}

\clearpage
\selectlanguage{english}

\begin{center}
\titleEN{AI-Generated English Translation}\medskip

\titleEN{Agentic AI for Management of Open RAN User-Centric Cell-Free Network}\medskip

Marcin Hoffmann$^{1,2}$

\medskip

\begin{minipage}[t]{0.6\textwidth}
\small $^{1}$ Poznan University of Technology, Poznań, \href{mailto:email}{marcin.hoffmann@put.poznan.pl} \\
\small $^{2}$ Rimedo Labs, Poznań \\
\end{minipage}
\end{center}

\medskip

\begin{multicols}{2}
\noindent
\abstractEN{One of the main challenges in implementing an Open RAN user-centric cell-free network is the proper formation of serving clusters. This problem can be effectively addressed by the proposed AI agent based on a large language model. Simulation results show that the proposed approach enables more than a 2\,$\times$ increase in the median of the user-throughput distribution compared to a state-of-the-art network-centric architecture.\footnote[1]{This work was carried out within the PRELUDIUM project funded by the National Science Centre (Poland), grant no. 2022/45/N/ST7/01930.}}
\medskip

\noindent
\keywordsEN{AI, 6G, User-Centric Cell-Free Architecture, Massive MIMO, Large Language Models.}

\section{Introduction}
One of the key techniques that can significantly improve the spectral efficiency of sixth-generation (6G) mobile networks is the so-called user-centric cell-free architecture~\cite{yang2021}. In this architecture, a user can be simultaneously served by multiple synchronized base stations (often referred to as access points), typically equipped with M-MIMO (Massive Multiple-Input Multiple-Output) antenna arrays. This architecture reduces disparities in the received power across the network, which is particularly important for users located at comparable distances from several base stations~\cite{ngo2015}. Moreover, coordinated beamforming can be used to mitigate interference.

From the practical-deployment viewpoint, the Open RAN (Open Radio Access Network) concept has strong potential for user-centric networks~\cite{Ranjbar2022,Hoffmann_2024_infocom}. It assumes the so-called 7.2 split, enabling the base-station functionality to be separated into a virtualized O-CU (Open RAN Centralized Unit), responsible, among others, for radio resource management; a virtualized O-DU (Open RAN Distributed Unit), implementing higher-PHY and MAC functions; and a physical O-RU (Open RAN Radio Unit), implementing lower-PHY functions and radio transmission. In user-centric networks, a single virtualized O-CU/O-DU instance can coordinate transmissions from many O-RUs, effectively acting as a CPU (Central Processing Unit). Additionally, thanks to open interfaces and the RIC (RAN Intelligent Controller), Open RAN enables deploying dedicated network-optimization algorithms, e.g., based on AI (Artificial Intelligence), including LLM-based AI agents~\cite{Wu2025}.

Due to limited computational resources at the O-CU/O-DU and O-RU sides, one of the main challenges in user-centric network management is serving-cluster formation~\cite{ammar2022}. The goal is to determine the set of O-RUs that should serve a given user. Because the number of possible combinations is large, this is a non-trivial task. If done improperly, it can even degrade network performance. For example, if too many O-RUs serve a user, distant O-RUs may have a marginal impact on throughput while wasting radio resources. On the other hand, too few O-RUs may not provide sufficient received-power gain or interference coordination. Therefore, using AI methods---especially language models that can interact with the environment and interpret results similarly to human-driven network configuration, but continuously and automatically---is a promising direction.

In this paper, I propose an AI agent for managing an Open RAN user-centric network. Due to the large action space, I adopt a hierarchical approach: an AI agent implemented as an rApp tunes the parameters of a low-complexity serving-cluster-formation algorithm implemented directly in the O-CU/O-DU based on the current user locations. The proposed AI agent maximizes the median of the user-throughput distribution and, unlike MAB (Multi-Armed Bandit) methods, leverages historical knowledge to reduce exploration time. The agent is validated using an advanced user-centric network simulator with 3D ray-tracing, time-frequency-space resource allocation, MCS (Modulation and Coding Schemes), and ZF (Zero-Forcing) precoding~\cite{hoffmann2024}.

\section{AI Agent for User-Centric Network Management} \label{sec:ai_agent_en}
\begin{figure}[H]
\centering
\includegraphics[trim={11cm 1.5cm 11cm 1.5cm},clip,width=0.49\textwidth]{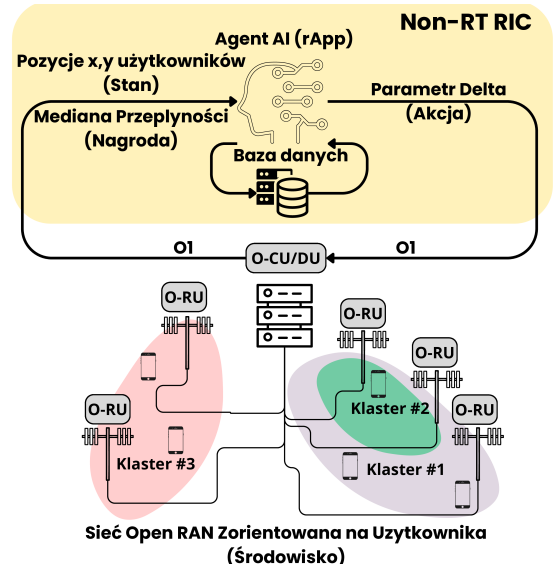}
\caption{AI agent for managing serving-cluster formation in a user-centric network.}
\label{fig:system_en}
\end{figure}

This paper proposes an AI agent responsible for downlink serving-cluster formation in an Open RAN user-centric network, as shown in Fig.~\ref{fig:system_en}. In this network, multiple M-MIMO equipped O-RUs are connected to a virtualized O-CU/O-DU acting as the CPU. Besides radio resource allocation and coordinated transmission to users from physical O-RUs, a key task of the O-CU/O-DU is to establish serving clusters. This requires configuring the O-RUs in a cluster to enable, e.g., user-data exchange, transmission synchronization, and sharing channel-state information. A particularly challenging case is when serving clusters require coordination between two O-CU/O-DU instances, i.e., network fragments controlled by different CPUs. Before that is possible, it must be determined which users should be served by which O-RUs---i.e., serving clusters must be formed.

In the proposed management scheme, we assume that the O-CU/O-DU contains a built-in low-complexity cluster-formation algorithm suitable for use cases requiring reaction times on the order of milliseconds. Such an algorithm was proposed in~\cite{tugfe2021cellfree}. The serving cluster of user $k$ (denoted by $\mathcal{M}_k$) is formed by the minimum number of O-RUs $N_k$, selected among all $L$ O-RUs (indexed by $l$), with the largest transmit power to the user $p_{k,l}$, such that
\begin{equation}\label{eq:serving_cluster_formulation_en}
\frac{\sum_{l \in \mathcal{M}_k}p_{k,l}}{\sum_{l=1}^{\mathrm{L}}p_{k,l}} \geq \Delta,
\end{equation}
where $\Delta$ is an algorithm parameter in the range $[0,1]$. Values of $\Delta$ close to $0$ correspond to a network-centric architecture (a user served by the single O-RU providing the highest received power), while $\Delta=1$ means that a user is served by all O-RUs within reach. Hence, the key challenge is selecting $\Delta$ appropriately depending on the current user spatial distribution.

To this end, I propose an AI agent which, due to its relatively long response time (on the order of a few seconds) and higher computational requirements, is implemented as an rApp in the Non-RT RIC (see Fig.~\ref{fig:system_en}). The agent-environment interaction follows a reinforcement-learning loop and uses the O1 interface to retrieve network data and update $\Delta$. In each cycle, the agent identifies the state, takes an action using historical knowledge stored in a database, and observes the reward. The loop elements are defined as follows:
\begin{itemize}
    \item \textbf{State:} a matrix of user positions expressed as X,Y coordinates.
    \item \textbf{Action:} selecting the parameter $\Delta$ from the range $[0,1]$.
    \item \textbf{Reward:} the median of the user-throughput distribution.
\end{itemize}
The proposed AI agent is based on a large language model (LLM). Its action-selection logic, given the current state and historical data, is described textually in:
\begin{itemize}
    \item \textbf{System prompt:} defines the situational context and the agent behavior. In the proposed system prompt, the agent is informed that it manages a user-centric network consisting of $L$ O-RUs. Next, the definition of $\Delta$ (range and interpretation) is provided. Finally, the agent is told that user positions (state) and historical data (median throughput observed for given positions and $\Delta$) will be provided as inputs. Based on that, the agent should return the value of $\Delta$ (action) that maximizes the throughput median (reward).
    \item \textbf{User prompt:} the direct query to the agent containing the inputs. For the proposed agent, the user prompt provides the current user locations (state) and historical data enabling the agent to infer which $\Delta$ to choose based on previously observed rewards. Moreover, the agent can use historical data to generalize to previously unvisited states without re-running an independent learning process for each new user-location pattern.
\end{itemize}

\section{Simulation Results} \label{sec:results_en}
The proposed AI agent for serving-cluster formation in an Open RAN user-centric network was evaluated using an advanced simulator proposed by the author in prior work~\cite{hoffmann2024}. The simulator models a user-centric M-MIMO OFDMA network. It uses a realistic radio-channel model based on 3D ray-tracing (Wireless InSite) and a modular processing chain including serving-cluster formation, a dedicated radio-resource-allocation algorithm, ZF precoding executed independently at each O-RU, and 15 modulation-and-coding schemes. Within each O-RU, the transmit power per resource block is equally split among users scheduled on that block via spatial multiplexing.

For the simulation experiments, the proposed AI agent uses the Qwen 3.5-32b language model deployed on a local workstation via Ollama. The remaining simulation parameters are listed in Table~\ref{tab1_en}.

\begin{table}[H]
\centering
\caption{Simulation parameters}
\label{tab1_en}
\medskip
\begin{tabularx}{1.0\columnwidth}{p{0.65\linewidth} p{0.17\linewidth} }
\toprule
\textit{Parameter} & \textit{Value}\\ \midrule
Language model & Qwen 3.5 32b \\ \hline
Carrier frequency & 3.6~GHz \\ \hline
Number of resource blocks & 69 \\ \hline
Number of O-CU/O-DU & 1 \\ \hline
Number of O-RU radio units & 1 macro, 5 micro \\ \hline
Antenna height---macro O-RU & 45~m\\ \hline
Antenna height---micro O-RU & 6~m\\ \hline
Transmit power---macro O-RU & $46$~dBm \\ \hline
Transmit power---micro O-RU & $30$~dBm \\ \hline
Number of antennas---macro O-RU & 128\\ \hline
Number of antennas---micro O-RU & 32 \\ \hline
Simulation time (slots) & 5000 \\ \hline
Number of users & 20 \\ \hline
Wireless InSite---number of reflections & 15 \\ \hline
Wireless InSite---number of diffractions & 1\\\bottomrule
\end{tabularx}
\end{table}

In the simulation study, I compared the proposed AI agent to: a network-centric architecture (denoted as 1 O-RU in the plots), a user-centric architecture where all O-RUs serve all users (denoted as 6 O-RU), and selecting $\Delta$ using a contextual multi-armed bandit (MAB) algorithm~\cite{Hoffmann_2024_krit}. The methods were compared across 5 independent states, i.e., independently drawn user-location layouts. Additionally, because LLMs are probabilistic and their outputs can vary between calls, 10 independent simulation runs were performed for the proposed AI agent using the same sequence of states.

\begin{figure}[H]
\centering
\includegraphics[width=0.49\textwidth]{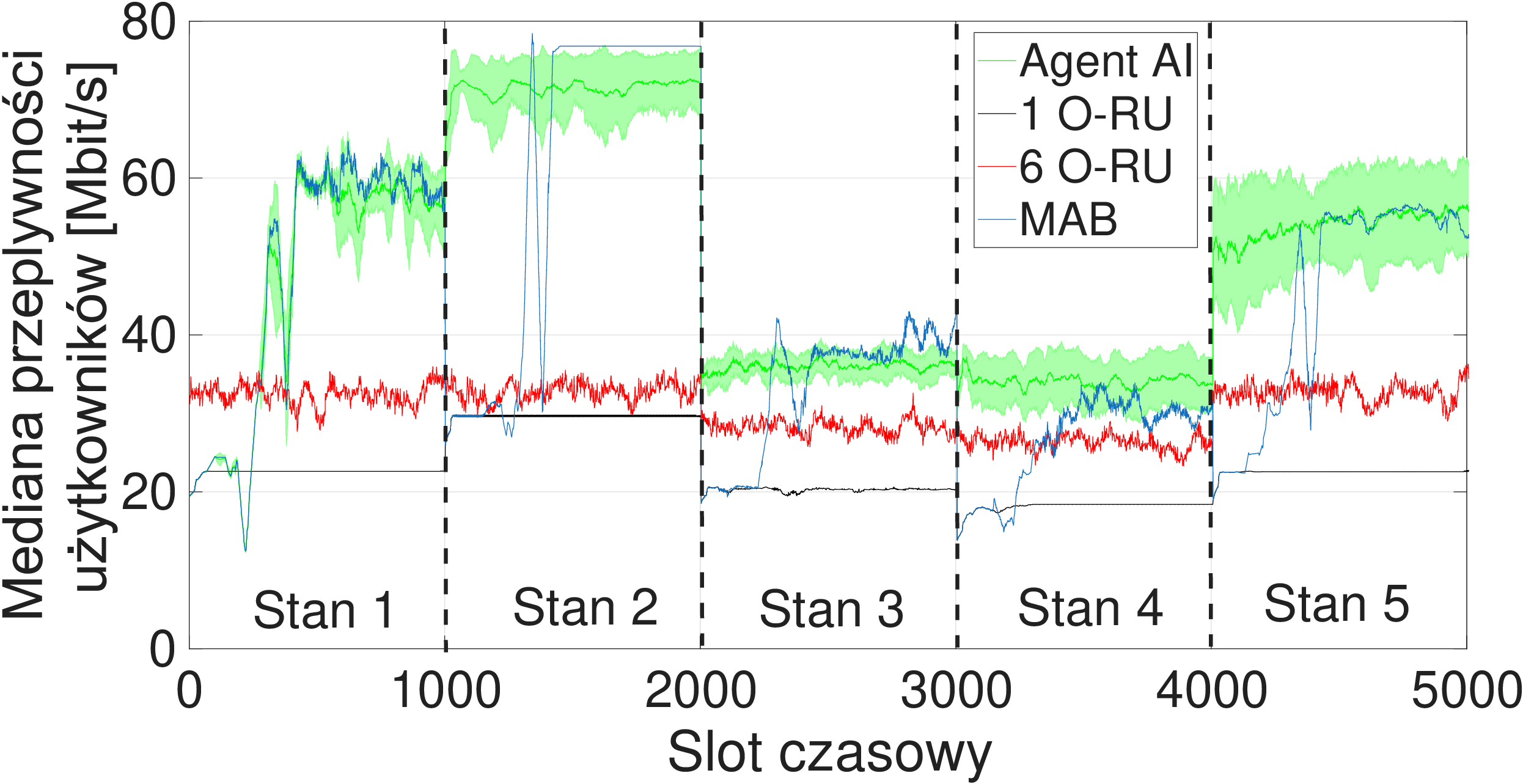}
\caption{Median user throughput versus time. For the AI agent, the mean and standard deviation over 10 independent simulation runs are shown.}
\label{fig:learning_time_en}
\end{figure}

The resulting time evolution of the throughput median is shown in Fig.~\ref{fig:learning_time_en}. First, in the considered states, the network-centric architecture (1 O-RU) achieves the lowest median throughput in each state: approximately 20--30~Mbit/s. In contrast, the extreme user-centric configuration where each user is served by all O-RUs yields only a modest improvement to roughly 25--35~Mbit/s. A substantial gain can be achieved by tuning $\Delta$ between 0 and 1. For MAB, selecting $\Delta$ is performed independently within each state and achieves medians in the range from 35~Mbit/s up to about 70~Mbit/s (State 2). On the one hand, independent per-state optimization may find a slightly better value in individual cases; on the other hand, it requires spending about one third of the time exploring all $\Delta$ values, including those that degrade performance. Finally, the proposed AI agent performs a full learning process only in State 1 and then reuses accumulated knowledge, making only minor adjustments. In most states, the mean median throughput achieved by the AI agent is comparable to MAB, but (except State 1) without repeating the full learning process each time. The throughput median fluctuates with a standard deviation of a few Mbit/s across independent runs.

\begin{figure}[H]
\centering
\includegraphics[width=0.49\textwidth]{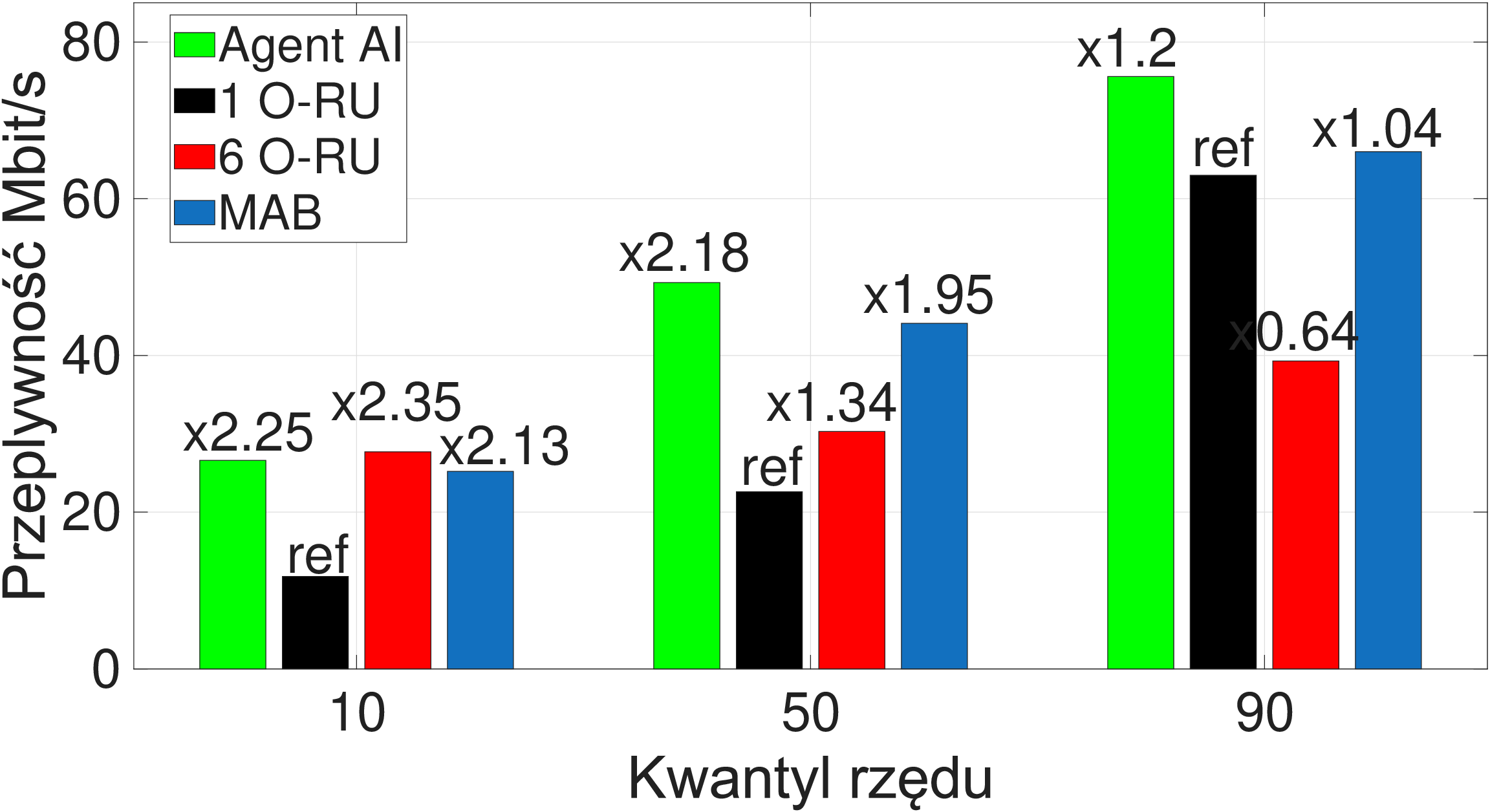}
\caption{Comparison of the 10th, 50th, and 90th percentiles of the user-throughput distribution over the entire simulation.}
\label{fig:rate_bar_en}
\end{figure}

To compare the AI agent and MAB more directly, Fig.~\ref{fig:rate_bar_en} presents the throughput median over the entire simulation experiment, together with the 10th and 90th percentiles, representing users with the worst and best propagation conditions, respectively. The network-centric result (1 O-RU) is used as the reference. For the user-centric architecture where all users are served by all O-RUs, we observe a 1.34$\,\times$ gain in median throughput and a 2.35$\,\times$ gain for worst-case users. For these users, interference coordination and increased received power are crucial. However, this comes at the cost of the best-case users, whose throughputs drop to 64\% of the reference, likely due to transmit-power sharing among a large number of (often distant) users.

The results also highlight the benefits of dynamically adapting $\Delta$, for both the proposed AI agent and MAB. In terms of the throughput median, the proposed AI agent improves the network-centric baseline by 2.18$\,\times$, while MAB yields 1.95$\,\times$ (about a 12\% gain in favor of the AI agent). This is primarily due to the AI agent's ability to infer a suitable $\Delta$ in States 2--5 from limited historical data (e.g., from State 1) without running an independent learning process each time, as required by MAB. Similar trends are observed for the worst-case users: the AI agent achieves a 2.25$\,\times$ gain versus 2.13$\,\times$ for MAB. Finally, for best-case users, the AI agent improves throughput by about 20\% relative to MAB, which provides only about a 4\% improvement.

\section{Conclusions} \label{sec:conclusions_en}
Serving-cluster formation is one of the key challenges for 6G user-centric networks. This paper proposes an LLM-based AI agent to address this problem. Unlike MAB methods, the AI agent can perform human-like reasoning based on historical data, which makes it possible to maintain a high median user throughput by applying only minor adjustments as user locations change. As a result, the proposed AI agent improves the median user throughput by about 12\% compared to the MAB solution. A future research direction is to compare the performance achieved by the proposed AI agent across different language models, including both smaller open models and large commercial cloud-hosted models such as ChatGPT or Gemini.

\end{multicols}
\end{document}